\documentclass[amsmath,amssymb,aps,prl,twocolumn,showpacs,superscriptaddress]{revtex4-1}
\usepackage[pdftex]{color,graphicx}
\pdfoutput=1

\newcommand{\ii}{i}

\newcommand{\eq}[1]{(\ref{eq:#1})}

\begin{document}

\title{Inter-mode reactive coupling induced by waveguide-resonator interaction}

\author{Mher Ghulinyan}
\affiliation{Centre for Materials and Microsystems, Fondazione Bruno Kessler, I-38123 Povo, Italy}

\author{Fernando Ramiro Manzano}
\affiliation{Department of Physics, Nanoscience Laboratory, University of Trento, I-38123 Povo, Italy}

\author{Nikola Prtljaga}
\email{Present address: Department of Physics and Astronomy, University of Sheffield, S3 7RH, United Kingdom }
\affiliation{Department of Physics, Nanoscience Laboratory, University of Trento, I-38123 Povo, Italy}

\author{Martino Bernard}
\affiliation{Centre for Materials and Microsystems, Fondazione Bruno Kessler, I-38123 Povo, Italy}
\affiliation{Department of Physics, Nanoscience Laboratory, University of Trento, I-38123 Povo, Italy}

\author{Lorenzo Pavesi}
\affiliation{Department of Physics, Nanoscience Laboratory, University of Trento, I-38123 Povo, Italy}

\author{Georg Pucker}
\affiliation{Centre for Materials and Microsystems, Fondazione Bruno Kessler, I-38123 Povo, Italy}

\author{Iacopo Carusotto}
\affiliation{INO-CNR BEC Center and Department of Physics, University of Trento, I-38123 Povo, Italy}

\begin{abstract}
We report on a joint theoretical and experimental study of an integrated photonic device consisting of a single mode waveguide vertically coupled to a disk-shaped microresonator. Starting from the general theory of open systems, we show how the presence of a neighboring waveguide induces reactive inter-mode coupling in the resonator, analogous to an off-diagonal Lamb shift from atomic physics. Observable consequences of this coupling manifest as peculiar Fano lineshapes in the waveguide transmission spectra. The theoretical predictions are validated by full vectorial 3D finite element numerical simulations and are confirmed by the experiments.
\end{abstract}
%\date{\today}

\maketitle

The study of the consequences of coupling a physical system to an environment constitutes the central problem in the theory of open systems~\cite{breuer}. This coupling, on one hand, allows the system to dissipate energy through active decay channels. On the other hand, its reactive component leads to a shift of energy levels and oscillation frequencies of the system. Most celebrated examples of this physics involve an atom coupled to the bath of electromagnetic modes~\cite{CCT4}, namely, the (dissipative) spontaneous emission of photons from an excited state~\cite{SERydberg,yablo,Vos} and the (reactive) Lamb shift of transition frequencies~\cite{lamb,lamb2,lamb3}.

Pioneering experimental studies in late 1970's~\cite{cpt-pisa} showed that destructive interference of different decay paths, leading to the same final continuum, can suppress absorption by a multilevel atom via the so-called Coherent Population Trapping (CPT)~\cite{arimondo_review} and Electromagnetically Induced Transparency (EIT)~\cite{EITexp1,EITreview} mechanisms. While originally these phenomena were discovered in the atomic physics context, a continuous interest has been devoted to analogous effects in solid-state systems~\cite{EITsolidstate}, photonic devices~\cite{matsko,lipson,21,20,22,EIT_THZ,oeHuang}, and, very recently, optomechanical systems \cite{OMIT}. Though in most experiments only the dissipative features are affected by the interference, the theory predicts that a similar phenomenon should also occur on the reactive coupling side~\cite{breuer}.

In photonics, the presence of a waveguide in the vicinity of a resonator activates new radiative decay channels for the resonator modes via emission of light into the waveguide mode~\cite{photonicsbook,photonicsbook2,yariv}. The corresponding reactive effect is a shift of the resonator mode frequencies, which can be interpreted as the photonic analogue of the atomic Lamb shift. In this Letter, we report on a joint theoretical and experimental study of a photonic device in which pairs of modes of very similar frequencies are coupled simultaneously to the same waveguide mode. Both the dissipative and the reactive couplings of the cavity modes to the waveguide turn out to be modulated by interference phenomena between the two modes, which can be summarized as environment-induced inter-mode coupling -- a sort of {\em off-diagonal Lamb shift} $\Delta_{ij}$ in the atomic analogy. In the experiments, these coupling terms are responsible for peculiar Fano interference lineshapes in the transmission spectra of single resonators.

\begin{figure}[t!]
\centering
\includegraphics[width=8.5cm]{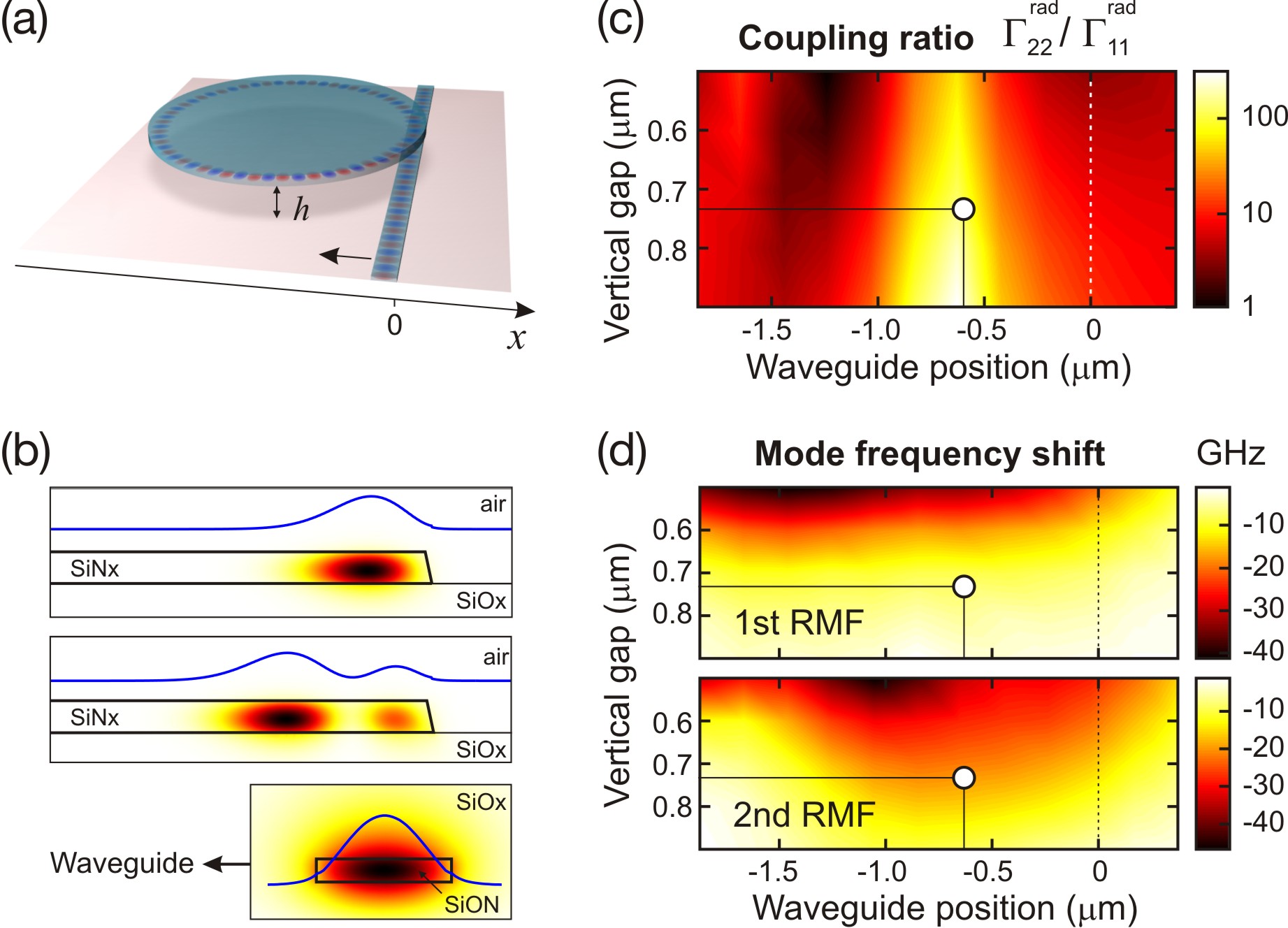}
\caption{(Color online) (a) A sketch of the microphotonic device. (b) The intensity profile of the first and second radial mode families (RMF) of the resonator (top and middle panels) and of the waveguide mode (bottom panel). Blue curves show the cuts of the intensity profile, and the labels indicate the different materials. Panels (c-d): Results of {\em ab initio} numerical calculations for (c) the radiative decay rate ratio $\Gamma_{22}^{\rm rad}/\Gamma_{11}^{\rm rad}$ and (d) the frequency shifts of the first (top) and the second (bottom) radial family modes as a function of the waveguide position. The open circles indicate the waveguide position for the fabricated 40~$\mu$m-diameter resonator.}
\label{fig:fig1}
\end{figure}

The system under consideration consists of a thin microdisk resonator vertically coupled to an integrated single-mode waveguide located below the disk (Fig.~\ref{fig:fig1}(a)). In contrast to the traditional lateral coupling geometry, where typically only the most external first radial mode family (RMF) experiences an appreciable coupling to the waveguide, the vertical coupling geometry allows for an independent lateral and vertical positioning of the waveguide, permitting thus to freely tune the coupling to the different mode families, in particular to the more internal ones~\cite{mherIEEE,verticup}. Since these latter typically have lower intrinsic quality factors, the vertical coupling geometry is crucial to our experiments, as it allows for several RMFs to be simultaneously close to critical coupling and, therefore, visible in transmission spectra.

This remarkable tunability is illustrated in Fig.\ref{fig:fig1}(c,d). An {\em ab initio} finite element numerical simulation of the 3D vectorial Maxwell equations, including the detailed geometrical shape and material composition of the resonator-waveguide system, is used to obtain the frequencies and damping rates of the different eigenmodes of the electromagnetic wave equation \cite{SM}. At this stage, we have focussed on a pair of spectrally distinct modes, in order to minimize the contribution from mode-coupling terms. The ratio of the decay rates (Fig.\ref{fig:fig1}(c)) is proportional to the relative intensity of the waveguide coupling to the lowest two RMFs of the resonator: as expected, its value is the largest when the lateral position of the waveguide matches the main lobe of the the second RMF (Fig.\ref{fig:fig1}(b) middle panel). The photonic analogue of the atomic Lamb shift for (independent) cavity modes is illustrated in Fig.\ref{fig:fig1}(d): the shifts $\Delta_{11}$ and $\Delta_{22}$ of the two modes from their {\em bare} frequencies are plotted as a function of the waveguide position. While in the atomic case the calculation of the Lamb shift, originating from photon emission/reabsorption processes, requires sophisticated techniques and a careful handling of UV divergences~\cite{CCT4}, in the photonic case one typically has a red-shift of all modes when a generic dielectric material is approached to a resonator~\cite{photonicsbook}.

\paragraph{Theoretical model --}
\label{sec:theory}

The transmission of a waveguide, $T(\omega)$, coupled to the resonator can be described by generalizing the input-output theory of optical cavities~\cite{Walls} to the multi-mode case. In the present two-mode case, the equation of motion for the field amplitudes $\alpha_{j=1,2}$ can be written as
\begin{multline}
 \ii \frac{d\alpha_j}{dt}=\left[\omega_j^o + \Delta_{jj} -\ii \frac{\gamma_j^{\rm nr}+\Gamma^{\rm rad}_{jj}}{2}\right]\alpha_j + \\
 +\left(\Delta_{12} - \ii \frac{\Gamma^{\rm rad}_{12}}{2}\right)\alpha_{3-j} +\bar{g}_j E_{\rm inc}(t) \label{eq:alpha1}
\end{multline}
In the absence of the waveguide, the two modes oscillate independently from each other at a bare frequency $\omega_j^o$ and have an intrinsic, non-radiative decay rate $\gamma_j^{\rm nr}$. The incident field, which propagates along the waveguide and drives the resonator, is described in the last term in Eq.~\eq{alpha1}. The coupling amplitude of the driven waveguide mode to the $j=1,2$ resonator mode is quantified by the $\bar{g}_j$ coefficients. In the following, we focus on a monochromatic excitation with $E_{\rm inc}(t)=E_{\rm inc}\,e^{-\ii\omega_{\rm inc}t}$.

The effect of the waveguide on the cavity mode oscillation is included in the motion equation Eq.~\eq{alpha1} via the Hermitian $\Gamma^{\rm rad}$ and $\Delta$ matrices, for which formal application of the theory of open systems within the Markov approximation~\cite{breuer} provides the general expression
\begin{equation}
\Delta_{jl}+\ii\,\frac{\Gamma^{\rm rad}_{jl}}{2}=\int\!\frac{dK}{2\pi}\,\sum_{\beta} \frac{g^*_{\beta,j}(K) \,g_{\beta,l}(K)}{\omega_{\rm inc}-\Omega_\beta(K)-\ii 0^+},
%\\ \simeq [\bar \Delta+\ii\bar \Gamma^{\rm rad}/2] g_j g_k^*:
\label{eq:K}
\end{equation}
in terms of the coupling amplitude $g_{\beta,j}(K)$ of the $j$th resonator mode to that of the waveguide of longitudinal wavevector $K$, mode index $\beta$, and frequency $\Omega_\beta(K)$.
For single-mode waveguides, $\Gamma^{\rm rad}$ is determined by the single propagating mode for which $\Omega_{\beta}(K)=\omega_{\rm inc}$ and $g_{\beta,j}(K)=\bar{g}_j\,$. This imposes that the $\Gamma^{\rm rad}_{12}$ coefficient, typically responsible for EIT-like interference effects in the atomic context, is related to the radiative linewidths $\Gamma_{jj}^{\rm rad}$ by $\Gamma^{\rm rad}_{12}=\sqrt{\Gamma^{\rm rad}_{11}\Gamma^{\rm rad}_{22}}$.

Even though a quantitative estimation of $\Delta$ using Eq.~\eq{K} is in most cases impractical as it involves a sum over all (both guided and non-guided) waveguide modes, this equation provides an intuitive picture of the underlying process: the diagonal and off-diagonal terms originate from the virtual emission of a photon from a resonator mode and its immediate recapture by the same or another mode, respectively. From a qualitative point of view, while the diagonal terms are typically $\Delta_{jj}<0$, we are unable to invoke any general argument to determine the non-diagonal $\Delta_{12}$. A similar inter-mode coupling term was mentioned in~\cite{20} starting from a coupled-mode approach. Here we will show how a real $\Delta_{12}>0$ is needed to reproduce the experimental data and we will point out some unexpected features due to this term.

\begin{figure}[t!]
\centering
\includegraphics[width=8.5cm]{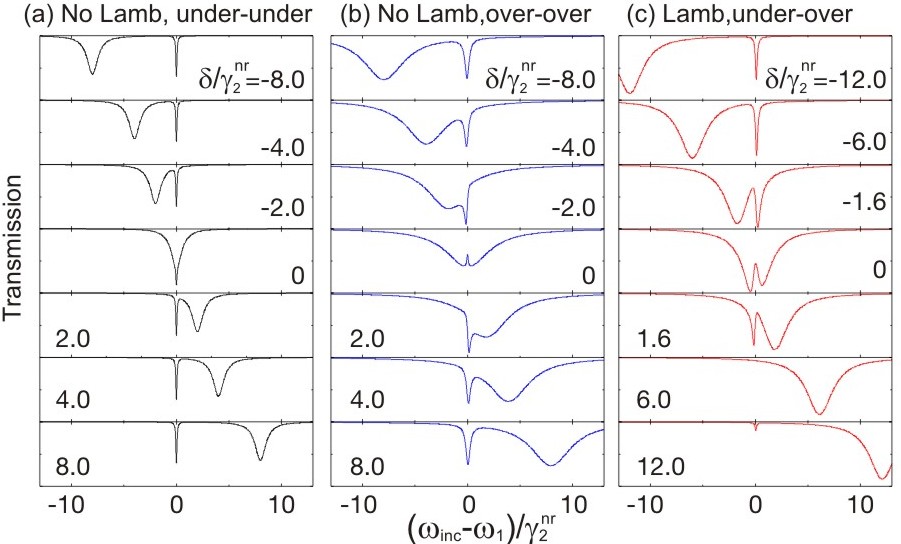}
\caption{(Color online) Analytically calculated transmission spectra in different regimes and for different detunings $\delta=\omega_2-\omega_1$. (a) Both modes are undercoupled, $\Gamma^{\rm rad}_{11,22}/\gamma^{\rm nr}_{1,2}=0.25$ and $\Delta_{12}=0$. (b) Both modes are overcoupled, $\Gamma^{\rm rad}_{11,22}/\gamma^{\rm nr}_{1,2}=4$ and $\Delta_{12}=0$. (c) Narrow (broad) mode is undercoupled (overcoupled), $\Gamma^{\rm rad}_{11}/\gamma^{\rm nr}_{1}=0.125$ ($\Gamma^{\rm rad}_{22}/\gamma^{\rm nr}_{2}=2$) and $\Delta_{12}/\gamma_2^{\rm nr}=0.6$.
The ratio $\gamma^{\rm nr}_1/\gamma^{\rm nr}_2=0.1$ (a-b) and $0.16$ (c).
}
\label{fig:theor}
\end{figure}

In our model, the waveguide transmission reads
$T(\omega_{\rm inc})=|{E_{\rm tr}}/{E_{\rm inc}}|^2= |1-\ii\,\rho\,{\sum_{j=1,2} \bar{g}^*_j \bar{\alpha}_j}/{E_{\rm inc}}|^2$, in terms of the stationary solution $\bar{\alpha}_j$ of the motion equations Eq.~\eq{alpha1} and the density of states $\rho=|dK/d\Omega|$ in the waveguide. When the waveguide is effectively coupled to one resonator mode only, a typical resonant transmission dip is recovered: under-, critical-, and over-coupling regimes are found depending on whether $\Gamma^{\rm rad}_{11}$ is lower, equal or larger than $\gamma_{1}^{\rm nr}$~\cite{photonicsbook,photonicsbook2,yariv}.

The novel and much richer phenomenology that occurs in the two-mode case is illustrated in Fig.~\ref{fig:theor}. Interesting features manifest clearly when both $j=1,2$ resonator modes are close to criticality $\gamma_{j}^{\rm nr}\approx \Gamma^{\rm rad}_{jj}$. In the left column of Fig.~\ref{fig:theor}, we show a case where both modes are slightly under-coupled $\Gamma^{\rm rad}_{jj}\lesssim\gamma^{\rm nr}_{j}$ and the off-diagonal reactive coupling vanishes $\Delta_{12}=0$. Each mode then manifests as a transmission dip in the spectrum centered at a frequency $\omega_j=\omega_j^o+\Delta_{jj}$ that includes the diagonal shift $\Delta_{jj}$. Note that the the first RMF is much narrower than the other since $\gamma_{1}^{\rm nr} \ll \gamma_{2}^{\rm nr}$ . Comparing the different rows of the figure, we notice that scanning the relative detuning of the two modes $\delta=\omega_2-\omega_1$ results in a simple, interference-free superposition of the two dips. Even in this simplest case, a correct inclusion of $\Gamma^{\rm rad}_{12}$ is, however, essential to avoid the appearance of nonphysical features in the calculations, such as $T(\omega)>1$.

The central column shows the case of slightly over-coupled modes $ \Gamma^{\rm rad}_{jj}\gtrsim \gamma^{\rm nr}_{j}$, still with $\Delta_{12}=0$. Now, marked interference features start to appear due to the off-diagonal dissipative coupling $\Gamma^{\rm rad}_{12}$ and the doublets of peaks acquire a complicated structure. In particular, the narrow dip, normally visible at $\omega_1$ (1st and 7th rows), is replaced by a complex Fano-like lineshape~\cite{fano1,CCT4} (3rd and 5th) for moderate detunings, and even reverses its sign into a transmitting EIT feature in the resonant $\delta=0$ case (4th row). Experimental observations of this physics were recently reported in~\cite{21,22,20}.

Finally, the dramatic effect of the off-diagonal reactive coupling $\Delta_{12}\neq0$ is shown in Fig.~\ref{fig:theor}(c). As most visible general feature, the spectrum is no longer symmetric under a change in the sign of $\delta$, and the spectral feature due to the narrow mode is clearly visible than one would expect given its deep under-coupling condition. With respect to the $\Delta_{12}=0$ case shown in Fig.~\ref{fig:theor}(b), the narrow Fano feature has a reversed sign for moderate detunings (3rd and 5th rows). Furthermore, it is suppressed in a finite detuning range (6th row). An analytical explanation of this unexpected effect is given in \cite{SM} in terms of the destructive interference of the direct excitation of mode 1 from the waveguide and its two-step excitation via mode 2 by the off-diagonal terms of $\Delta$ and $\Gamma$. The two paths almost cancel out around $\delta\simeq \Delta_{12} \sqrt{\Gamma^{\rm rad}_{22}/\Gamma^{\rm rad}_{11}}$.

\begin{figure}[t!]
\centering
\includegraphics[width=6cm]{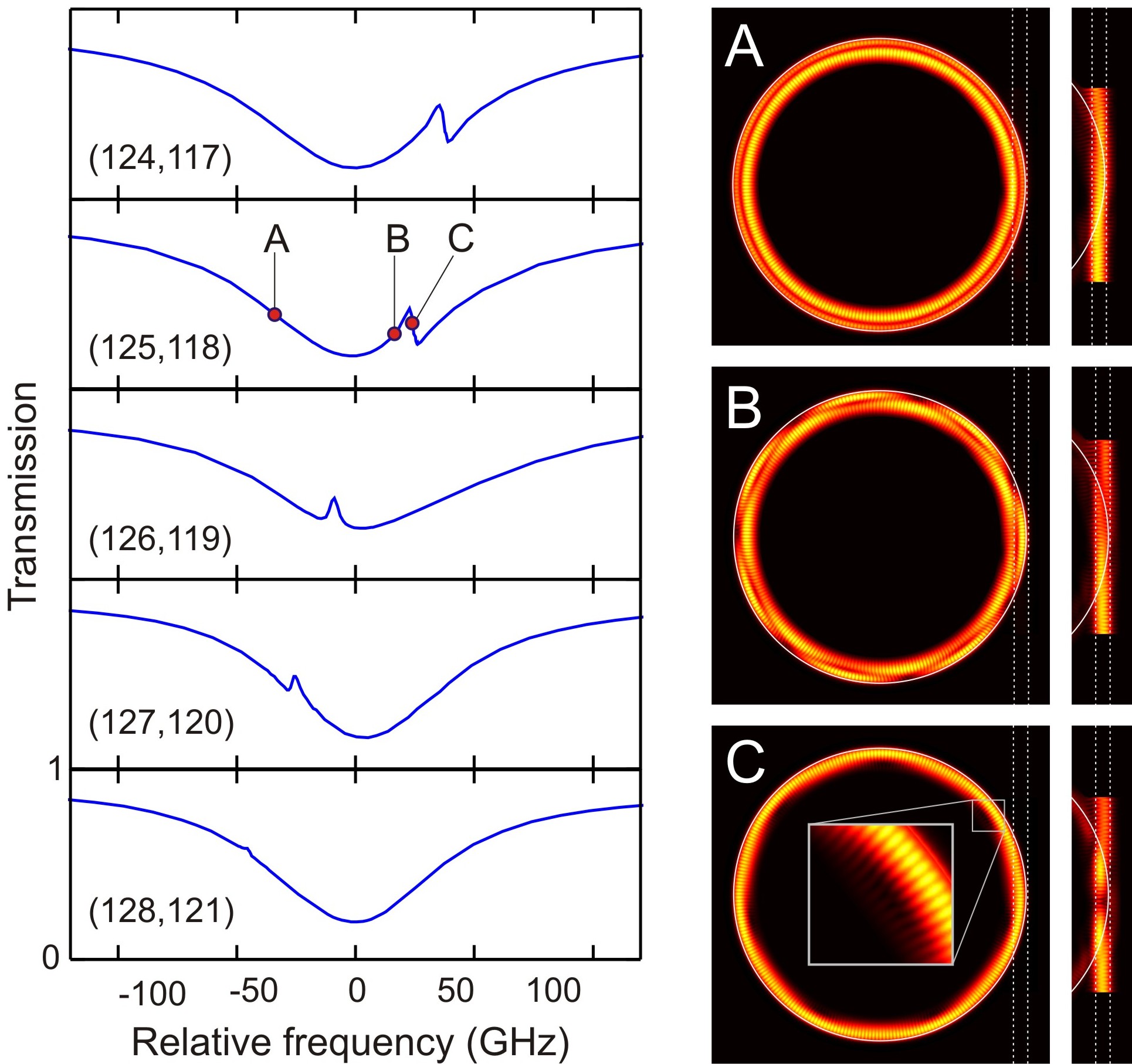}
\caption{(Color online) (Left) Numerically calculated spectra for different detunings between the two radial family modes. The azimuthal mode order of the two resonances is reported in each graph. (Right) planar cuts of the intensity profile within the resonator and within the waveguide at the frequencies indicated as A, B, C in the left panel.
}
\label{fig:sim}
\end{figure}

\begin{figure*}[t]
\centering
\includegraphics[width=13.2cm]{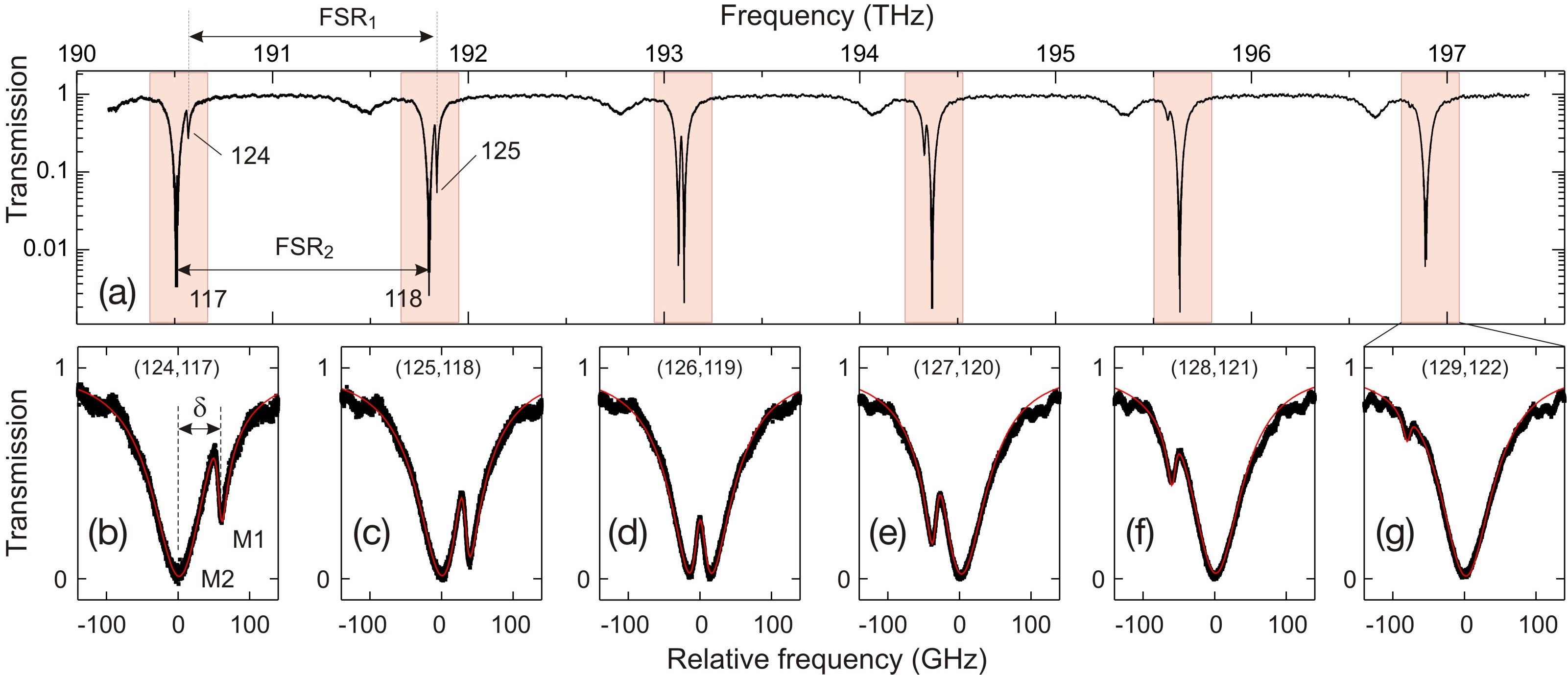}
\caption{(Color online) (a) Transmission spectra of a 40~$\mu$m diameter microresonator as a function of the absolute incident frequency. (b-g) Blow-ups of the regions marked as gray in (a). In each panel, the relative frequency is measured from the broader second family resonance.
Red lines show fits to the spectra using the analytical model.
}
\label{fig:exp40}
\end{figure*}

\paragraph{Numerical simulations --}
The analytical predictions for the transmission have been validated through {\em ab initio} finite element numerical simulations \cite{SM}. The slightly diverse free spectral range of the different RMFs allowed us to scan the relative detuning of the interfering modes by looking at pairs of quasi-resonant modes with different azimuthal quantum numbers. Examples of spectra are shown in the left panels of Fig.~\ref{fig:sim}. The qualitative agreement with the predictions of Eq.~\eq{K} in the $\Delta_{12}>0$ regime is remarkable: the Fano-like feature is clearly visible with the correct sign for generic detunings (1st to 4th rows) and disappears completely in a well-defined range of $\delta$'s (lowest row).
The three (A,B,C) panels on the right column of Fig.~\ref{fig:sim} show horizontal cuts of the field intensities in the resonator and the waveguide at three different incident frequencies across the Fano-like feature. While the excitation at the A (C) point is concentrated in the second (first) mode, interference between the two modes is responsible for the snaky shape of the intra-cavity intensity distribution at the intermediate point B. As expected, the number of spatial oscillations is determined by the difference in azimuthal quantum numbers of the two resonator modes.

\begin{figure} [t!]
\centering
\includegraphics[width=\columnwidth]{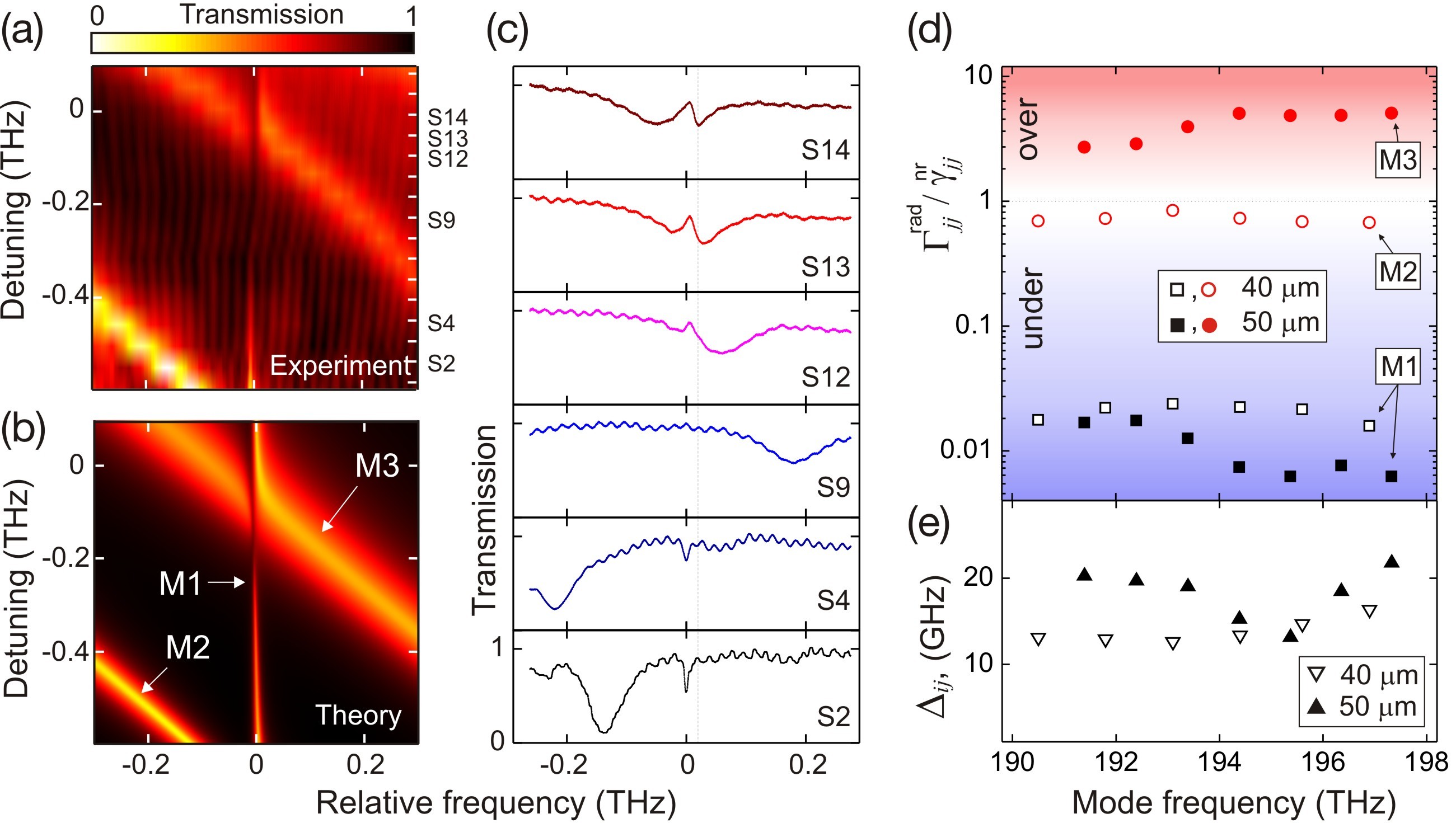}
\caption{(Color online) (a) Colormap merging 21 experimental transmission spectra (indicated as S1--S21) for a $50\,\mu$m resonator. On each row, the relative frequency is measured from the narrow mode frequency.
(b) Analytic prediction for $T(\omega)$ using a three-mode extension of the model with optimized global parameters. (c) Selected examples of spectra.
(d) System parameters obtained by independently fitting each experimental spectrum with the analytical model.
}
\label{fig:exp50}
\end{figure}

\paragraph{Experimental observations --}
We have validated our findings experimentally by looking at pairs of quasi-resonant modes originating from different radial families in microdisk resonators coupled vertically to dielectric waveguides. The experimental transmission spectrum through the waveguide for a microdisk of radius $R=40\,\mu$m is shown in the upper panel of Fig.~\ref{fig:exp40} \cite{SM}. The spatial position of the waveguide is indicated by open dots in Fig.\ref{fig:fig1}(c-d). The transmission spectrum consists of a sequence of doublets originating from the first (narrow features) and second (broader features) RMFs, which have slightly diverse free spectral range. This last permits to sweep one RMF across the other as the azimuthal order of the underlying modes is increased, and the doublets structure is correspondingly changed. In the bottom panels Fig.~\ref{fig:exp40}(b-g), zoomed views of the different doublets are shown: to facilitate comparison, in each of these panels the central frequency is located at the broader second family resonance (i.e. at $\omega_2=\omega_2^o+\Delta_{22}$ in the analytical model). These spectra are in excellent qualitative agreement with the predictions of the numerical simulations shown in Fig.~\ref{fig:sim}: the Fano-like feature has the correct sign and is visible for generic detunings exception made for a small range of values where it completely disappears [panel (g)]. Moreover, the experimental data are successfully fitted by the analytical model (red curves in Fig.~\ref{fig:exp40}(b-g)).

The generality of our observations has been confirmed by repeating the experiment on a larger $R=50~\mu$m resonator in which the Fano interference takes place between the first  and the third  RMFs. The measured transmission spectra are shown in Fig.~\ref{fig:exp50}(a,c) for different values of the relative detuning of the quasi-resonant pairs of modes. The crossing of the two families again leads to Fano interference profiles, and the narrow feature disappears in a specific range of detunings (spectrum S9). Furthermore, the experimental results successfully compared to the prediction of the analytical model, generalized to three modes (Fig.~\ref{fig:exp50}(b)).

Finally, Fig.~\ref{fig:exp50}(d-e) summarizes the fits parameters for both $40~\mu$m and $50~\mu$m resonators. Despite the total independence of the fit procedures performed on each spectrum, a smooth dependence of all fit parameters on the azimuthal mode number is observed. As expected, the scan of the azimuthal quantum number varies the mode detuning without affecting the other system parameters. From the top graph we notice that in both cases the first family modes are undercoupled to the waveguide, while the second and third family modes are very close to critical coupling. As stated in the theoretical section, this combination of couplings is crucial for a neat observation of the Fano feature. Finally, Fig.~\ref{fig:exp50}(e) shows that the fitted value of the off-diagonal reactive coupling $\Delta_{12}$ is always around $15$~GHz.

To summarize, in this work we have reported a joint theoretical and experimental study of a microdisk resonator vertically coupled to a single-mode waveguide. The importance of the inter-mode reactive coupling due to the neighboring waveguide is revealed and characterized from the peculiar Fano lineshapes manifesting in transmission spectra. In addition to its intrinsic interest for photonics, our study provides a simple model where to study a fundamental feature of the theory of open systems, namely the possibility of environment-mediated couplings -- the off-diagonal photonic Lamb shift -- between different modes of a system.

We acknowledge financial support Autonomous Province of Trento, Call {\em "Grandi Progetti 2012"}, project {\em "On silicon chip quantum optics for quantum computing and secure communications - SiQuro"}. IC acknowledges partial support from ERC via the QGBE grant.

\end{document}